\documentclass[preprint]{aastex}

\begin{document}

\title{Chandra X-ray Observatory Study of the Orion Nebula Cluster 
and BN/KL Region}

\author{ 
Gordon Garmire\altaffilmark{1}, Eric D. Feigelson\altaffilmark{1}, 
Patrick Broos\altaffilmark{1}, Lynne A. Hillenbrand\altaffilmark{2}, 
Steven H. Pravdo\altaffilmark{3}, Leisa Townsley\altaffilmark{1}, 
Yohko Tsuboi\altaffilmark{1}
}

\altaffiltext{1}{Department of Astronomy \& Astrophysics, 
525 Davey Laboratory, Pennsylvania State University, University Park 
PA 16802}
\altaffiltext{2}{Department of Astronomy, MS 105-24, California 
Institute of Technology, Pasadena CA 91125}
\altaffiltext{3}{Jet Propulsion Laboratory, MS 306-438, 4800 Oak 
Grove Drive, Pasadena CA 91109}

\shorttitle{Chandra study of the Orion Nebula and BN/KL}
\shortauthors{G. Garmire et al.}

\begin{abstract} 

About 1000 X-ray emitting young pre-main sequence (PMS) stars
distributed in mass from $\sim$0.05 M$_\odot$ brown dwarfs to a $\sim
50$ M$_\odot$ O star are detected in an image of the Orion Nebula
obtained with the Advanced CCD Imaging Spectrometer on board the Chandra
X-ray Observatory.  This is the richest field of sources ever obtained
in X-ray astronomy.  Individual X-ray luminosities in the Orion Nebula
Cluster range from the sensitivity limit of $<2 \times 10^{28}$ erg
s$^{-1}$ to $\sim 10^{32}$ erg s$^{-1}$.  ACIS sources include $85-90$\%
of $V<20$ stars, plus a lower but substantial fraction of deeply
embedded stars with extinctions as high as $A_V \simeq 60$.

The relationships between X-ray and other PMS stellar properties suggest
that X-ray luminosity of lower-mass PMS stars depends more on mass, and
possibly stellar rotation, than on bolometric luminosity as widely
reported.  In a subsample of 17 unabsorbed stars with mass $\simeq 1$
M$_\odot$, X-ray luminosities are constant at a high level around $L_x
\simeq 2 \times 10^{30}$ erg s$^{-1}$ for the first $\simeq 2$ My while
descending the convective Hayashi track, but diverge during the $2-10$
My phase with X-ray emission plummeting in some stars but remaining high
in others.  This behavior is consistent with the distribution of X-ray
luminosities on the zero-age main sequence and with current theories of
their rotational history and magnetic dynamos.

The sources in the Becklin-Neugebauer/Kleinman-Low (BN/KL) region of
massive star formation are discussed in detail.  They include both
unabsorbed and embedded low-mass members of the Orion Nebula Cluster,
the luminous infrared Source n, and a class of sources without optical
or infrared counterparts that may be new magnetically active embedded
PMS stars.  Several X-ray sources are also variable radio emitters, an
association often seen in magnetically active PMS stars.  Faint X-ray
emission is seen close to, but apparently not coincident with the
Becklin-Neugebauer object.  Its nature is not clear.

\end{abstract}


\section{Introduction}

X-ray emission among pre-main sequence (PMS) stars with masses $M \la 2$
M$_\odot$ is known to be elevated $10^1$ to $>10^4$ times above typical
main sequence levels \citep{Feigelson99}.  High-amplitude X-ray
variability and hard spectra, supported by multiwavelength studies,
indicate the emission is in most cases due to solar-type magnetic flares
where plasma is heated to high temperatures by violent reconnection
events in magnetic loops.  However, the empirical relationships between
PMS X-ray emission and stellar age, mass, radius, and rotation are
complex \citep[e.g.,][]{Feigelson93} and their links to astrophysical
theory are not strong.  For example, if surface magnetic activity is a
manifestation of a simple rotation-driven magnetic dynamo in the stellar
interior, and if current theories about the evolution of angular momentum
in PMS stars hold \citep{Bouvier97}, then specific predictions about the
evolution of the X-ray luminosity function during the PMS phase can be
tested if sufficiently large samples were obtained.  X-rays are known to
be produced in high-mass star formation regions but little is known about
the structure or astrophysical origins of this emission
\citep{Churchwell99}.

The largest sample of PMS stars observable in a single field is the
Orion Nebula Cluster (ONC).  The OB members of the ONC illuminate the
Orion Nebula (= Messier 42), a blister H{\sc II} region at the near edge
of the Orion A giant molecular cloud.  The ONC is a nearby and one of
the densest PMS stellar clusters known with $\simeq 2000$ members
concentrated in a 1 pc ($8^{\prime}$) radius sphere with 80\% of the
stars younger than 1 My \citep{Hillenbrand97}.  It was the first star
forming region to be detected in the X-ray band \citep{Giacconi72} and
non-imaging studies soon found the X-ray emission is extended on scales
of a parsec or larger \cite{denBoggende78,Bradt79}.  Early explanations
for the Orion X-rays included winds from the massive Trapezium stars
colliding with each other or the molecular cloud, and hot coronae or
magnetic activity in lower mass T Tauri stars.  The Einstein
\citep{Ku79}, ROSAT \citep{Gagne95, Geier95, Alcala96} and ASCA
\citep{Yamauchi96} imaging X-ray observatories established that both the
Trapezium stars and many lower-mass T Tauri stars contribute to the
X-ray emission.

These studies could detect only a modest fraction of the ONC stars due
to the high stellar concentration and heavy absorption by molecular
material along the line of sight.  NASA's Chandra X-ray Observatory
launched on 23 July 1999 with its Advanced CCD Imaging Spectrometer
(ACIS) detector provides an unprecedented combination of capabilities
$-$ a wide spectral bandpass, $<1$\arcsec\/ spatial resolution, very low
cosmic ray and detector background levels, and moderate spectral
resolution at every pixel $-$ that make it particularly well-suited to
the study of crowded and absorbed clusters of faint X-ray sources.

After describing the data in \S 2, we discuss demographics of PMS stellar X-ray
emission (\S 3).  The next sections describe two of many possible
investigations of this rich dataset:  in \S 4 we examine the evolution of
magnetic activity in solar-mass stars, and in \S 5 we present the X-ray sources
from stars in the Becklin-Neugebauer/Kleinman-Low (BN/KL) star forming region
in OMC 1.  Later papers will provide a more comprehensive analysis of this and
additional observations of the Orion Nebula (Feigelson et al., in preparation).

\section{Observations and data analysis}

The ONC was observed with the ACIS-I array detector on board the Chandra
X-ray Observatory \cite{Weisskopf96} for 47.8 ks on 12 October 1999.
The array consists of four abutted charge-coupled devices (CCDs)
specially designed for X-ray studies (Garmire et al., in
preparation\footnote{Detailed descriptions of the ACIS instrument and
its operation can be found on-line at www.astro.psu.edu/xray/docs/sop
and www.astro.psu.edu/xray/docs/cal\_report.}).  Two chips from the
ACIS-S spectroscopic array were also turned on, but their data will not
be discussed here.  Starting with Level 1 processed event lists provided
by the pipeline processing at the Chandra X-ray Center, we removed
several types of CCD events likely dominated by particles rather than
imaged X-rays:  energies above 8 keV and below 0.2 keV; split events
except for ASCA grades 0, 2, 3, 4 and 6; cosmic ray events causing
stripes along amplifier nodes; hot columns and other low-quality flagged
events.  The aimpoint on chip I3 was $5^h 35^m 15.0^s$ $-5^\circ
23^\prime 20^{\prime\prime}$.

The resulting image (Fig.\ 1) shows hundreds of sources concentrated
around the bright Trapezium stars.  Several instrumental effects are
also evident such as the broadening of the point response function
off-axis due to the telescope optics, the 11\arcsec\/ gaps between the
CCD chips blurred by the satellite aspect dither, and the trailing of
events from the brightest sources due to events arriving during the chip
readout.  The brightest Trapezium sources also suffer photon pileup
which distorts their morphology, flux and spectrum.

\placefigure{Fig1}

An additional problem in the current dataset is the degradation of
charge transfer efficiency of the ACIS-I CCDs due to high dosage of
particle irradiation early in the Chandra mission \citep{Prigozhin00}.
Corrective measures were taken to arrest the degradation as soon as it
was discovered and data acquired after 29 January 2000 should have a
reduced problem due to changes in on-board operation of the detector to
mitigate charge transfer inefficiency.  Improved data processing
techniques have been developed to treat this problem \citep{Townsley00}
and are applied in our analysis in \S 5.  The results of other portions
of the study should not be seriously affected by the charge transfer
degradation.

The excellent point spread function produced by the Chandra mirrors and
the satellite aspect solution, combined with the very low background
rate in the detector, permits the detection of sources as weak as
$\simeq 7$ photons and the resolution of sources as closely spaced as
2\arcsec.  This high sensitivity and resolution is illustrated in Fig.\
2.  For a weakly absorbed ONC star with a typical PMS thermal spectrum
with energy $kT = 1$ keV \citep[see][]{Preibisch97}, a limit of 7
photons corresponds to a flux of $9 \times 10^{-16}$ erg s$^{-1}$
cm$^{-2}$ ($0.2-8$ keV) or, assuming the ONC distance is 450 pc, a
luminosity of $2 \times 10^{28}$ erg s$^{-1}$.  This is $\simeq 25$
times more sensitive than the ROSAT HRI study of the ONC \citep{Gagne95}
and $\simeq 10$ times more sensitive than the ROSAT PSPC observation
which suffered considerable source confusion \citep{Geier95}.  For
comparison, the emission of the contemporary Sun in the same X-ray band
ranges from $\sim 2 \times 10^{25}$ erg s$^{-1}$ in its quiet phase to
$\sim 1 \times 10^{28}$ erg s$^{-1}$ during its most violent flares
\citep{Peres00}.  Note, however, that the conversion between count rate
and luminosity will be much higher for stars which are heavily absorbed.
Accurate luminosities for the entire sample require spectral analysis
for each star, which is beyond the scope of the present paper.

\placefigure{Fig2}

To locate sources in a methodical fashion, we apply the source detection
program {\it wavdetect} based on a Mexican hat wavelet decomposition and
reconstruction of the image \citep{Freeman00}.  We set the significance
criterion at $1 \times 10^{-5}$, which permits a few false positive
detections but gives good sensitivity.  Wavelet scales between 1 and 16
pixels are used.  The procedure is nearly complete to a limit of 7
photons near the field center increasing to 10 photons towards the edges
of the field; many sources below these completeness limits are also
located.  A careful examination of the field by eye indicates the
wavelet detection algorithm is quite reliable.  The principal
adjustments to the source lists were the removal of some dozens of
spurious wavelet sources associated with the trails of the Trapezium
sources and the addition of 19 sources just resolved ($1-3$\arcsec
separation) from brighter sources.  Source detection was performed
separately in the soft ($0.2-2$ keV) and hard ($2-8$ keV) bands to
improve sensitivity to unembedded and embedded sources respectively,
after which the source lists were merged.

This procedure located 831 sources above the completeness limit and an
additional 142 likely sources below this limit (Table 1).  The sources should
be designated CXOONC hhmmss.s-ddmmss, where the acronym stands for `Chandra
X-ray Observatory Orion Nebula Cluster'.  Source count rates range from 0.1 to
$>400$ cts ks$^{-1}$ (the upper limit here is due to loss of events due to
photon pileup in $\theta^1$C Ori), giving a very high dynamic range.  This list
of 973 ACIS sources was cross-correlated with a large catalog of $\simeq 2500$
objects in the ONC and its vicinity including complete listings of 1578 stars
with visible magnitude $V<20$ \citep{Hillenbrand97} and 854 stars from K-band
surveys with limiting magnitude $K<18$ in the central region
\citep{Hillenbrand98, Hillenbrand00}.  We find that 860 of the 973 X-ray
sources spatially coincide within 1\arcsec$-$2\arcsec\/ of $V$- or $K$-band
cluster members.  Systematic and statistical astrometric errors are $<
1$\arcsec\/ (see \S 5).

\placetable{Table1}

In addition to the optical-infrared stars, most of the known compact
radio sources in the ONC region \citep[][and references
therein]{Felli93} are detected in the ACIS image.  These radio sources
are a heterogeneous mixture of ionized stellar winds or nonthermal
emission from both unobscured Trapezium O stars and embedded OB stars in
the BN/KL region, magnetically active low mass PMS members, and sources
without stellar counterparts such as partially ionized globules within
the giant Orion HII region \citep{Garay87, Felli93}.  We discuss some of
these sources in \S 5.

\section{Demographics of X-ray emission in the ONC}

Figure 2 illustrates several types of X-ray source identifications.  The
majority of the 973 X-ray sources are detected in both the soft and hard
X-ray bands and are associated with relatively bright stars seen in the
$V$ and $K$ surveys (sources with circle, square, $V$ and $K$).  Faint
unobscured stars may be found only in the soft X-ray band (circle, $V$
and $K$) while embedded stars are often seen only in the hard X-ray and
$K$ bands (square and $K$).  Many of the fainter and more heavily
absorbed ONC members, and a handful of unabsorbed stars, are not
detected in X-rays ($V$ and $K$ or $K$ alone).

One hundred and three ACIS sources found only in the hard band have no
optical or infrared counterpart and are distributed approximately evenly
across the ACIS field (square).  Some of these are likely extragalactic
sources, primarily active galactic nuclei, seen through the molecular
cloud.  Convolving the flux distribution of hard X-ray band source
counts at high-Galactic latitudes \citep[Ann Hornschemeier, private
communication; see][]{Hornschemeier00} with the distribution of hydrogen
column densities across the ACIS field (estimated from a
velocity-integrated C$^{18}$O map using standard conversions to total
column densities; John Bally, private communication) predicts $\sim 25$ 
hard extragalactic sources in the ACIS field.  We conclude that most of 
the unidentified sources are ONC stars, probably faint embedded late-type 
PMS stars (\S 5).

Nine hundred and thirty four of the $V<20$ stars are sufficiently
well-characterized to be placed on the Hertzsprung-Russell (HR) diagram
from which, together with theoretical evolutionary tracks
\citep{Dantona97}, stellar masses and ages can be estimated
\citep{Hillenbrand97}.  Figure 3 shows this HR diagram with symbols
coded for detection and non-detection in the ACIS image.  The detection
rate in the sample is 91\% for stars with masses $M > 0.3$ M$_\odot$ and
75\% for stars with $0.1<M<0.3$ M$_\odot$.  The Chandra sensitivity of
$\sim 2 \times 10^{28}$ erg s$^{-1}$ is thus sufficient to detect nearly
all optically selected ONC stars with $V<20$.

\placefigure{Fig3}

However, the $V$-band sample in the HR diagram systematically excludes
the most obscured and least massive stars, particularly brown dwarfs,
which are more effectively found at near-infrared wavelengths.  Figure 4
compares the ACIS sources to the deep $H$ and $K$ band samples
\citep{Hillenbrand98, Hillenbrand00}.  A strong relationship between
X-ray detectability and $K$ magnitude is present.  Over 50\% of cluster
members with $M > 1$ M$_\odot$ are detected, even if they suffered
absorption as high as $A_V \simeq 60$, equivalent to a hydrogen column
density $N_H \simeq 1 \times 10^{23}$ cm$^{-2}$.  Stars between $0.1 < M
< 1$ M$_\odot$ are generally detected only if $A_V \leq 10$, and
substellar objects with $M < 0.08$ M$_\odot$ are rarely detected for any
$A_V$.  Astrophysical interpretation of these results is not simple
because mass, luminosity, radius, reddening and the effects of
circumstellar disks are intertwined across the color-magnitude diagram.
To remove the effect of obscuration, we restrict our examination to
low-$A_V$ stars having infrared colors $(H-K) < 0.5$.  Again a strong
relationship between ACIS detection fraction and $K$ magnitude is found:
the detection fraction declines from about 85\% for $K < 11$ to about
50\% for $11 \leq K < 13$, about 35\% for $13 \leq K < 15$, and $< 8$\%
for $K > 15$.

\placefigure{Fig4}

A few of the 973 ACIS sources coincide with faint sources with inferred $M
\simeq 0.05$ M$_\odot$, assuming an age of 1 My (Figures 3 and 4).  These
candidate PMS brown dwarfs are PC 116, PC 215, H 5096, HC 114 and HC 123
\citep{Prosser94, Hillenbrand00}.  Two of these have spectral type around
M6$-$M7.  Taken at face value, these brown dwarf candidates roughly doubles the
number of PMS candidate brown dwarfs detected in the X-ray band at levels around
$10^{28}$ erg s$^{-1}$ with the ROSAT satellite \citep{Neuhauser98,
Neuhauser99}.  However, because these ONC stars reside close to the $M = 0.08$
M$_\odot$ line, infrared excesses from circumstellar disks may be present,
photometric variability may occur, and theoretical models are uncertain, they
may in fact evolve into hydrogen-burning stars.  The strongest result from the
ACIS ONC data is that great majority of PMS brown dwarfs found by Hillenbrand \&
Carpenter (2000) are not X-ray emitters with $L_x \ga 2 \times 10^{28}$ erg
s$^{-1}$, our detection limit, in agreement with the ROSAT studies.

\section{Evolution of X-ray emission in solar mass stars}

Among solar-type {\it main sequence} stars, it is well-established that
X-ray emission arises from magnetic activity at the stellar surface,
where the magnetic fields are generated by a magnetic dynamo in the
stellar interior driven by rotation \citep{Rosner85}.  Thus, main
sequence stars show a correlation between $L_x$ and surface rotational
velocity but little dependence between $L_x$ and other stellar
properties such as bolometric luminosity $L_{bol}$ or mass $M$.  The
situation for {\it pre-main sequence} stars has been more confusing:
despite considerable evidence that their X-ray emission is related to
high levels of surface magnetic activity, $L_x$ has been found to be
strongly correlated with both $L_{bol}$ and $M$
\citep[e.g.,][]{Walter81, Feigelson93, Casanova95, Neuhauser95,
Feigelson99}.  Furthermore, as PMS stars evolve downward in the
Hertzsprung-Russell diagram during their early convective phases, a
dependence on $L_{bol}$ may mean that X-ray luminosity decreases with
age $t$ during PMS evolution.  $L_x$ may also scale with stellar radius
(recall $R \propto L_{bol}^{1/2}$ along vertical convective tracks) if
reconnecting magnetic fields cover the entire surface of the star.  A
weak relation between $L_x$ and stellar rotation is found in some, but
not all, PMS samples.

These relationships have never been convincingly explained.  If
$L_x-L_{bol}$, $L_x-M$ or $L_x-t$ correlations represent physical links,
then PMS magnetic activity may not be based on the standard dynamo but
rather on more exotic mechanisms.  Such mechanisms include surface
activity arising from magnetic fields inherited from the star formation
processes rather than generated within the PMS stars, or violent
reconnection events in magnetic fields connecting the star to the
circumstellar disk \citep{Feigelson99}.

To untangle a part of this confusion, we consider only a narrow range of
stellar masses around 1 M$_\odot$ to remove mass as a confounding
variable.  Figure 5 shows the plot of ACIS count rate $vs.$\/ bolometric
luminosity for a complete sample of 17 ONC stars with $0.8 < M < 1.2$
M$_\odot$ and low obscuration ($A_V < 3$).  A well-defined sample of
Pleiads in the same mass range observed with the ROSAT satellite,
selected to have $0.52 \leq (B-V) \leq 0.74$, is shown for comparison
\citep{Stauffer94, Micela96}.  These studies established that rotational
velocity is the dominant contributor to the wide dispersion in Pleiad
X-ray emission.

Rather than a simple dependence between $L_x$ and $L_{bol}$, the X-ray
emission of ONC solar-mass stars in Fig.\ 5 might be divided into two
epochs.  During the first $\sim 2$ My as the bolometric luminosity fades
by a factor of $\sim 20$, X-ray luminosities appear high and constant
around $L_x \sim 2 \times 10^{30}$ erg s$^{-1}$.  This means the X-ray
surface flux ($L_x/4 \pi R^2$) increases as the star descends the
Hayashi track, as found in ROSAT studies \citep{Neuhauser95, Kastner97}.
But as the stars age from $\sim 2$ to $\sim 10$ My and enter the
radiative track, some stars retain their high X-ray levels while in
others it drops by a factor of $\sim 50$ or more.  The sample is small,
so this pattern is not established with confidence.

\placefigure{Fig5}

This behavior is in accord with current views of a solar-type magnetic
dynamo origin for PMS X-ray emission where rotation regulates the level
of magnetic activity at the stellar surface, and where rotation in turn
is regulated by a magnetic coupling between the star and its
circumstellar disk.  In this model, the wide dispersion in $L_x$ among
older PMS stars reflects the wide dispersion in stellar rotation rates
which is expected from different durations of star-disk coupling
\citep{Bouvier97}.  This same explanation accounts for the wide
dispersion of $L_x$ linked to stellar rotation found in zero-age main
sequence solar-mass Pleiades stars.

This dynamo interpretation of PMS X-ray emission can be directly tested
using the sample shown in Fig.\ 5:  older ONC stars with high $L_x$
(JW198, JW601 and JW738) should show faster rotational velocities in
high-resolution optical spectra than those with low $L_x$ (JW278, JW662
and JW868).  If this prediction of the standard magnetic dynamo model is
not validated in this and similar subsamples of the ONC, then non-solar
models for magnetic flaring in T Tauri stars must be considered.

\section{X-rays from the embedded BN/KL region}

A cluster of embedded stars, including the $L_{bol} \sim 10^4$ L$_\odot$
Becklin-Neugebauer (BN) object lie near the density peak of the massive
OMC-1 molecular cloud about 1\arcmin\/ northwest of the Trapezium
\citep{Genzel89}.  BN lies in a group of several $L_{bol} \sim 10^3$
L$_\odot$ massive stars \citep[][and references therein]{Gezari98},
three of which (BN, IRc2, and infrared Source n) produce compact thermal
radio sources (ultracompact HII regions and/or stellar winds) and/or
maser outflows.  This group of massive young stars illuminates the
infrared Kleinman-Low (KL) nebula and powers strong shocked molecular
outflows in the region \citep[][and references therein]{Stolovy98}.
BN/KL lies in the densest and most chemically rich region in the entire
Orion giant molecular cloud complex, and is the nearest and best studied
example of $\simeq 20$ known molecular hot cores which are thelikely
sites of current massive star formation \citep{Kurtz00}.  The region is
extremely complex and not fully understood after decades of intensive
multiwavelength investigation.

To obtain a reliable detailed understanding of the X-ray emission in
this crowded region, the ACIS data were specially processed for this
analysis.  First, we apply the software corrector for charge transfer
inefficiency (\S 2) described by Townsley et al.\ (2000).  This restores
much of the degradation of event energies and grade distributions in a
spatially-dependent fashion across the CCD chip.  Second, source
detection proceeded as described in \S 2.  Third, source positions of
several bright X-ray sources were matched to positions of infrared
sources \citep[aligned to the Hipparcos frame through the 2MASS
survey;][]{Hillenbrand00} and VLA radio sources \citep{Felli93,
Menten95}.  A 0.5\arcsec\/ boresight correction was applied to the ACIS
field resulting in a systematic astrometric accuracy better than
$\pm$0.1\arcsec\/ (estimated 90\% confidence level).  Individual source
positions have estimated statistical uncertainties ranging from $\pm
0.1$\arcsec\/ for strong sources to $\pm 0.5$\arcsec\/ for weak sources
(estimated 90\% confidence).

Fourth, spectral analysis on the corrected data was performed with the
XSPEC software package \citep{Arnaud96}.  We present here the equivalent
hydrogen column density associated with line-of-sight absorption.  Due
to uncertainties in the detector spectral response and the complexities
of source spectra, we estimate the 90\% confidence intervals to be $\pm
0.5$ in log$N_H$.  Recall that log$N_H = 21.2$ cm$^{-2}$ corresponds to
$A_V = 1$ for normal elemental abundances and gas-to-dust ratios
\citep{Ryter96}.  Fifth, photon arrival times were tested for
variability using a Kolmogorov-Smirnov one-sample test against a model
of a constant source.  We report source variations significant at the $P
> 99.5$\% confidence level.

\placefigure{Fig6}

\placetable{Table2}

Figure 6 shows a full-resolution view of the ACIS field around BN/KL in
the $0.2-8$ keV band.  Unlike Figures 1 and 2, colors here represent the
photon energy where red represents hard photons and blue represents soft
unabsorbed photons.  Twenty-seven sources above the limit of 7 photons
(\S 2) found with wavelet detection and visual examination are
indicated; their X-ray properties and counterparts are given in Table 2
and the Table Notes.  Two sources may be poorly resolved doubles with
separations around 1.5\arcsec.  Fainter sources are likely present; for
example, a group of 5 photons around ($\Delta \alpha, \Delta
\delta$)=(12\arcsec,$-$16\arcsec) appears as a source in Table 1 and
coincides with an anonymous $K \simeq 18-19$ NICMOS star
\citep{Stolovy98}.

In Table 2, running source numbers shown in Figure 6 are provided in
column 1.  X-ray centroid positions in columns $2-3$ are in epoch J2000
from the ACIS image after boresight correction.  Column 4 gives ACIS
counts extracted from a 1.5\arcsec\/ circle in the $0.2-8$ keV band with
ASCA grades after correction for charge-transfer inefficiency.  A
background of 2 counts per source has been subtracted.  Column densities
rounded to the nearest dex(0.5) in column 5 are estimated from XSPEC
fits to X-ray pulse height distributions.  Column 6 provides offsets in
arcseconds between the ACIS source and its counterpart, based on K-band
\citep{Hillenbrand00} or radio positions \citep[][adjusted to the
astrometry of Menten \& Reid 1995]{Felli93}.  

We divide sources into three classes based on their counterparts:  `ONC'
ACIS sources are likely members of the Orion Nebula Cluster, generally
optically-bright lightly-absorbed stars; `IR' ACIS sources have infrared
but not optical counterparts and are likely embedded young stars; and
`X-ray' ACIS sources have no photospheric stellar counterparts.  Columns
$8-9$ of Table 2 indicate the optical counterparts and magnitudes (JW =
Jones \& Walker 1988, P = Prosser et al.\ 1994).  Columns $10-11$ give
near-infrared counterparts and K-band magnitudes (JW = Jones \& Walker 1988,
PC = Prosser et al.\ 1994, HC = Hillenbrand \&
Carpenter 2000), except for magnitudes of BN and Source n obtained from
Lonsdale et al.\ (1982).  The final column points to Table Notes on
individual sources where optical counterpart information is from
Hillenbrand (1997), infrared information is from Dougados et al.\
(1993), and radio information is from Felli et al.\ (1993), Menten \&
Reid (1995) and Menten (private communication).  X-ray variability and spectral properties are also
summarized in the Notes.

\noindent {\bf Orion Nebula Cluster members} \\ The majority of ACIS
sources in the vicinity of BN are relatively unobscured ($A_V \la 3$)
members of the ONC, as expected given the proximity ($\simeq 1$\arcmin\/
or $\simeq 0.1$ pc) from the Trapezium.  The X-ray stars are very young
with estimated ages below 1 Myr.  Most are lower mass stars with $M
\simeq 0.1-0.4$ M$_\odot$, although two X-ray bright 0.8 M$_\odot$ stars
are present.  The poorly studied late-K star JW 470 = CXOONC
J053515.35-052215.8 is the most magnetically active star of this class,
exhibiting the strongest and most variable X-ray emission.  The
identification of CXOONC J053515.07-052231.3 with the unobscured optical
star PC 086 is uncertain despite the 0.5\arcsec\/ spatial coincidence, as
the X-ray source appears heavily absorbed while the optical star is not.
An alternative possible counterpart is the 12.4$\mu$m source IRc14 lying
2.0\arcsec\/ from the X-ray location \citep{Gezari98}.

Two ACIS sources in this class, CXOONC J-53514.08-052237.1 =  JW 432 and
CXOONC J053514.66-052233.9 = JW 448, are radio sources.  These and other
radio sources discussed here are generally variable and are classified
as FOXES, Fluctuating Optical and X-ray Emitting Sources, by Garay
(1987).  They resemble the several dozen known weak-lined T Tauri stars
seen in nearby star forming regions exhibiting nonthermal
gyrosynchrotron radio emission and strong X-ray emission \citep{Andre96,
Feigelson99}.  These radio/X-ray ONC stars have probably lost their
interacting disks at a very young age, resulting in rapid rotation and
high levels of magnetic flaring.

\noindent {\bf Embedded infrared stars} \\ Six ACIS sources are
associated with embedded 2$\mu$m stars without visible counterparts.
Four of these are consistent with embedded but otherwise ordinary
low-mass ONC members.  CXOONC J053514.35-052232.9, however, coincides
with the luminous infrared Source n.  This $L_{bol} \simeq 10^2$
L$_\odot$ member of the BN/KL group lies at the center of the dense
molecular hot core and a H$_2$O maser outflow, and appears to be
associated with both a nonthermal double radio source and diffuse
thermal radio emission \citep{Menten95, Chandler97, Gezari98}.  The
unabsorbed X-ray luminosity of order $10^{30}-10^{31}$ erg s$^{-1}$ is
consistent with a magnetically active low-mass protostar or T Tauri star
or with an ordinary early-type B star.

Perhaps most remarkable among these sources is CXOONC J053514.06-052222.2 which
lies 1.1\arcsec\/ northwest of the BN object itself.  This offset appears
larger than the expected positional uncertainties, estimated at $\pm
0.5$\arcsec\/ at 90\% confidence.  But only $\simeq 18$ source photons are
present and we cannot be completely confident the offset is real.  One
possibility is that the X-ray source is unrelated to the BN object but arises
from an uncatalogued embedded ONC star; the energy distribution of the detected
X-ray photons suggests $A_V \sim 50$.  Another possibility is that the emission
arises from outflows from the BN object.  The photon distribution is consistent
with a source structure $\la 1$\arcsec\/ ($\simeq 500$ A.U.)  in extent.
Hydrodynamical calculations of ultra-compact HII regions with mass-loaded winds
from massive young stars pushing into a stationary dense medium indicate that,
although the emission is dominated by readily-absorbed ultraviolet photons,
some high-energy X-rays are produced and escape from the immediate environment
\citep[][see also Arthur et al.\ 1993]{Strickland98}.  There are also
possibilities of non-equilibrium shock processes giving rise to X-ray emission
at the interfaces between H{\sc II} regions and molecular material
\citep{Dorland87}.

\noindent {\bf X-ray sources without optical or infrared counterparts} \\ The
five ACIS sources in the `X-ray' class with no photospheric counterparts in the
deepest available optical, K-band and mid-infrared surveys of the region
\citep{Stolovy98, Hillenbrand00, Gezari98}.  All five sources are heavily
absorbed with log$N_H \simeq 23$ cm$^{-2}$ or roughly $A_V \sim 50$, two
exhibit short-timescale X-ray variability, and three are associated with
compact, likely variable radio sources \citep{Felli93}.  The two ACIS sources
not associated with radio sources have unusually hard X-ray spectra similar to
those seen in Class I protostars and T Tauri stars during powerful flares
\citep{Koyama96, Tsuboi98}.  CXC J053514.90-052225.7 = radio D is the most
extreme of these sources.  Its luminosity would be $L_x \simeq 10^{31}$ erg
s$^{-1}$ if it were unobscured.

Three explanations for these sources might be considered:  background
Galactic or extragalactic sources unrelated to the Orion star forming
region; X-ray luminous low-mass protostars \citep[one of the dozen known
X-ray emitting Class I protostars is a nonthermal radio variable][]
{Feigelson98}; or magnetically active low-mass radio-loud (i.e.\
weak-lined T Tauri) stars.  We estimate that only $0-1$ of these six
sources may be an extragalactic object seen through the molecular cloud
(\S 3).  Protostars are possible, but perhaps their disks or outflows
should have been detected in the extensive infrared and molecular
studies of the region.  We suggest that the third alternative is most
plausible.  The X-ray and radio properties are consistent with deeply
embedded low-mass PMS stars having unusually high $L_x/L_{bol}$ ratios
such that their photospheric emission falls below current infrared
sensitivity limits \citep[$K \simeq 19$;][]{Stolovy98}.

It is worth remarking on the absence of certain objects in BN/KL region.
The failure to detect the widely-discussed multiple infrared sources
comprising IRc2 \citep{Dougados93} could be due to their high level of
absorption \citep[$A_V \simeq 58$;][]{Gezari98}.  No extended emission
is evident in this portion of the ACIS image.  X-rays are thus not
detected from infrared-luminous dust condensations \citep[IRc3, IRc4,
IRc7, infrared Shell or Crescent;][]{Stolovy98}, or the molecular hot
core, plateau or H$_2$ outflows \citep[e.g.,][]{Wright96}.  A recent
supernova explosion in the region \citep{Kundt97} is clearly excluded by
the low background level seen in Figure 6.

Independent of the nature of the host stars, a clear astrophysical
implication of the X-ray emission from deeply embedded PMS stars in the
BN/KL region and the wider ONC field is that X-ray dissociation regions
\citep{Hollenbach97} will be present even in the densest molecular cloud
cores.  These are volumes around X-ray emitting PMS stars where X-ray
ionization dominates cosmic ray ionization.  X-ray irradiation on
molecular cloud interiors is predicted to have a variety of effects on
the dynamics, astrochemistry, and dust characteristics of molecular
material \citep{Glassgold00}, although little evidence for such effects
has yet been found.

\section{Conclusions}

A 48 ks Chandra ACIS image of the Orion Nebula in the $0.2-8$ keV band
with a limiting sensitivity of $\simeq 2 \times 10^{28}$ erg s$^{-1}$,
sub-arcsecond spatial resolution and astrometric precision reveals:

\begin{enumerate}

\item About one thousand X-ray sources are associated with magnetically
active pre-main sequence stars in the Orion Nebula Cluster.  Almost 90\%
of the visible stars with $V<20$ are detected across the entire
Hertzsprung-Russell diagram.  Higher mass stars are detected even when
subject to tens of magnitudes of absorption, while the lowest mass stars
and substellar objects are typically not detected.

\item The relationships between X-ray emission and other stellar
properties are complex and should be elucidated in a more complete
analysis of the ONC.  Examination of a well-defined subsample of solar
mass stars suggests a high constant X-ray luminosity around $2 \times
10^{30}$ erg s$^{-1}$, hence increasing surface flux, as stars descend
the convective Hayashi track followed by an increased dispersion in
X-ray luminosities as they enter the radiative track.  The suggested
interpretation, that this dispersion in magnetic activity is due to
different rotational evolutions of different stars, can be easily tested
with measurements of stellar rotational velocities.  These results may
be relevant to the evolution of magnetic activity in the early Sun.

\item A close examination of X-ray sources in the BN/KL region of OMC 1
shows a diversity of stellar sources:  foreground and embedded ONC
members; the luminous embedded infrared Source n; several sources not
appearing in sensitive optical and K-band surveys; and a faint, heavily
absorbed source apparently displaced 1\arcsec\/ from BN itself.  The
X-ray spectra indicate line-of-sight absorption ranging from $21 \la
\log N_H \la 23$ cm$^{-2}$ and one-fourth of the sources are variable
within the observation.  A considerable fraction of the sources are
associated with faint radio sources; this likely arises from nonthermal
gyrosynchrotron radiation that accompanies X-ray flaring induced by
magnetic reconnection.  The ACIS source associated with Source n may be
the first unambiguous (i.e.\ spatially resolved from other young stars)
X-ray detection of a massive young stellar object.  The X-ray source
near BN is enigmatic.  It is unclear whether it is a separate star or
whether the X-rays are produced in the outflow from this massive young
star.  The several sources without any photospheric optical or infrared
counterpart are likely new embedded, magnetically active stars in OMC 1.
The presence of deeply embedded X-ray sources in the BN/KL region
support the idea that ionizing X-rays produced by young stars will be
present within dense star-forming molecular cloud cores.

\end{enumerate}

\acknowledgements

We express our appreciation of the many scientists and engineers who
brought Chandra to fruition, in particular those at MIT, Penn State and
Lockheed-Martin who contributed to the ACIS instrument.  Ann
Hornschemeier (Penn State), Karl Menten (MPI-Radioastronomie) and
John Bally (Colorado) kindly provided results prior to publication, and 
the referee Ralph Neuh\"auser gave many useful comments.  This research 
was funded by NASA contract NAS8-38252 at Penn State.  The contributions of 
S.H.P.  were carried out at the Jet Propulsion Laboratory, California Institute
of Technology, under contract with the National Aeronautics and Space
Administration.

\newpage

\bibliography{aj-jour}

\newpage 

\begin{figure}
\epsscale{1.0}
\plotone{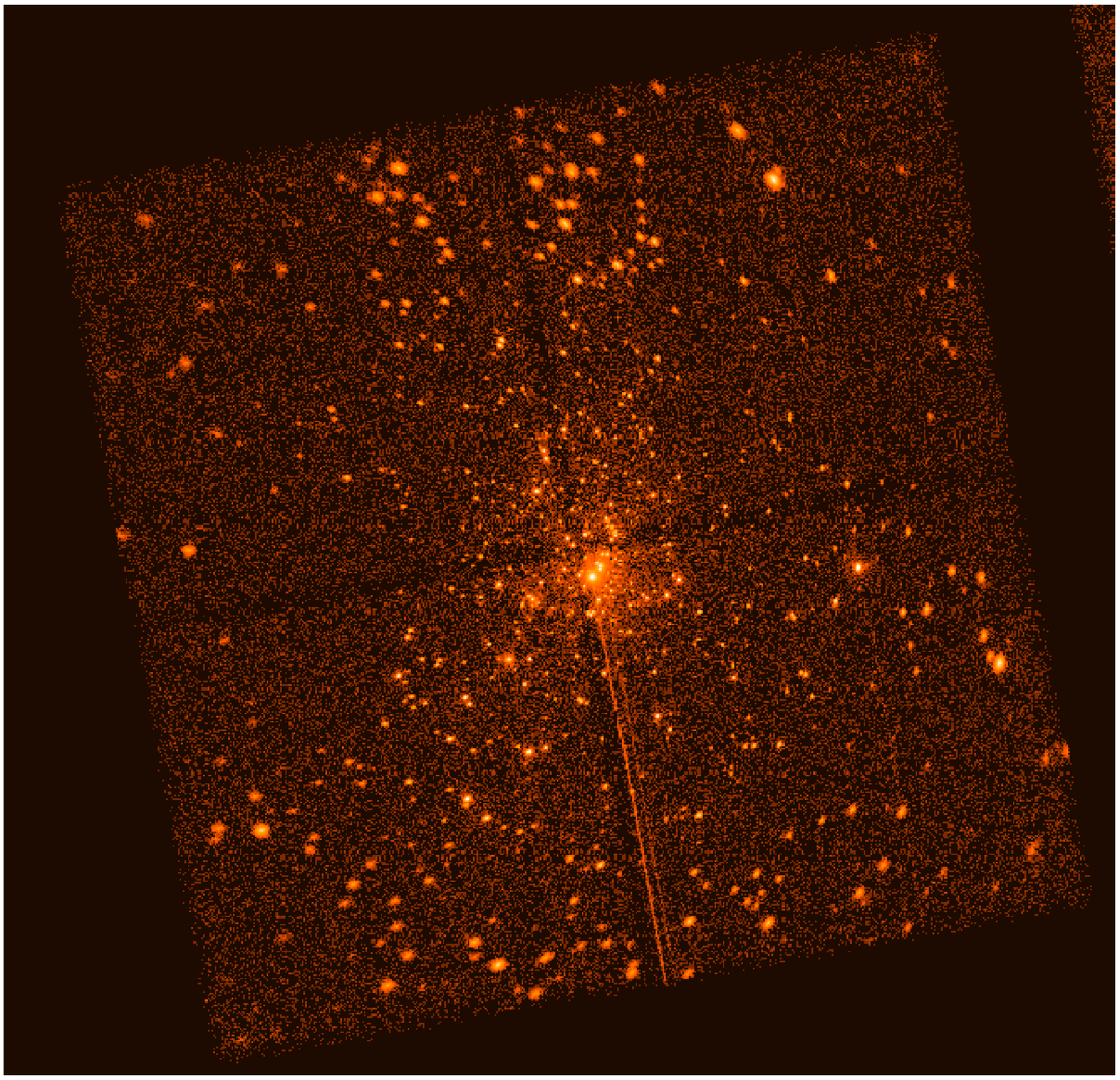}
\end{figure}

\figcaption[garmire.fig1.eps]{ACIS-I image of the Orion Nebula Cluster in the
$0.2-8$ keV band showing $\simeq 1000$ X-ray sources.  The
17\arcmin$\times$17\arcmin\/ array is shown here at reduced resolution
with 2\arcsec$\times$2\arcsec\/ pixels.   The massive O7 star
$\theta^1$C Ori is the brightest source near the center.  Here and in
Figure 2, colors are scaled to the logarithm of the number of events in
each pixel. Pixels with no events are black, those with individual
events are dark red, and those with multiple events are scaled
logarithmically.  North is at top and East is to the left.  
\label{Fig1}}

\begin{figure}
\epsscale{1.0}
\plotone{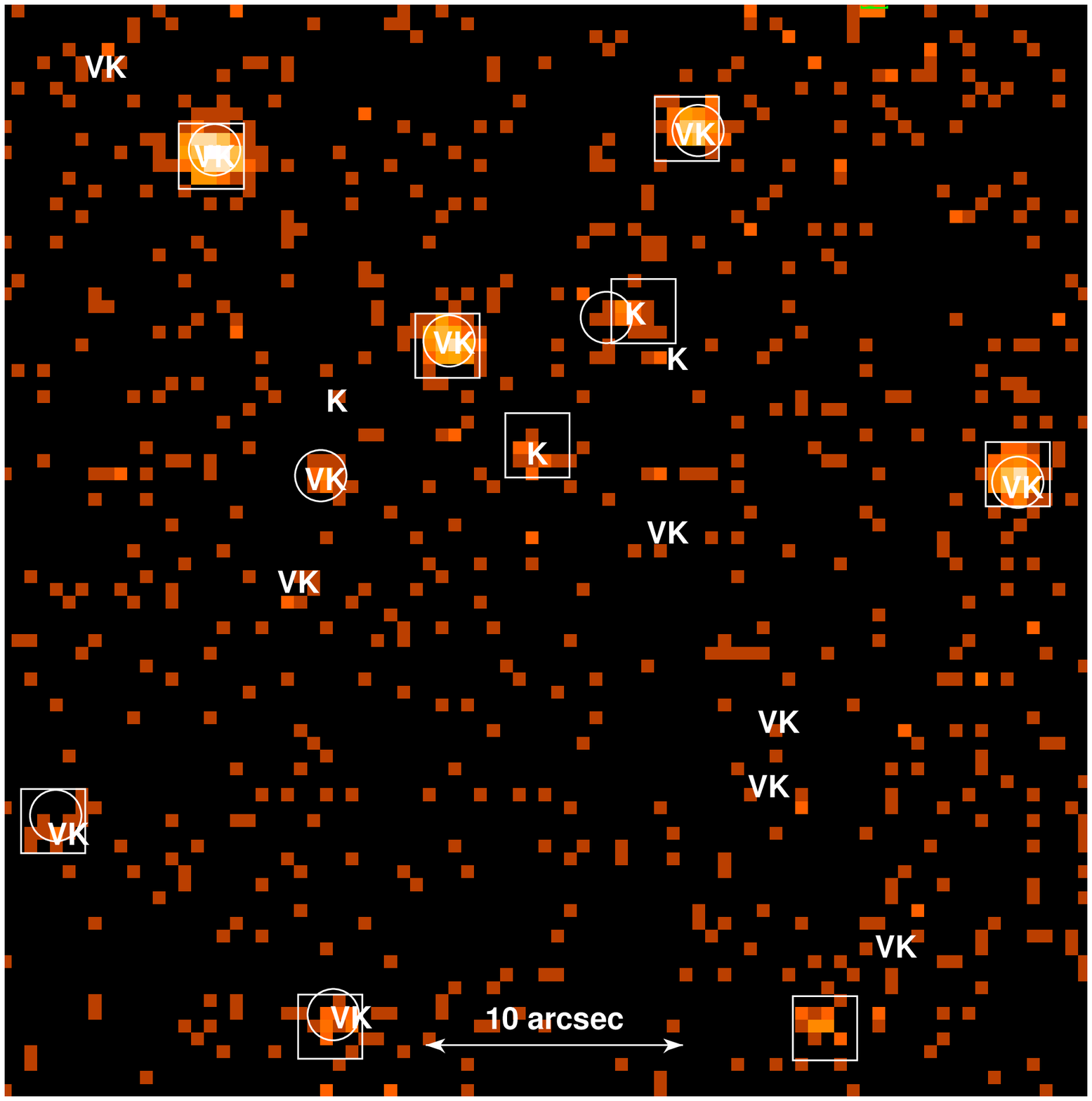}
\end{figure}

\figcaption[garmire.fig2.eps]{Expanded view of a typical small region at
full resolution with 0.5\arcsec$\times$0.5\arcsec\/ pixels.  The
background level is very low at $\simeq 0.08$ cts pixel$^{-1}$.  The
bright source near the western edge, Parenago 1959 ($V=16.5$ K star), was
the only source previously detected in this field (Gagn\'e et al.\ 1995).  White
circle and squares show ACIS sources located with the wavelet detection
algorithm in the soft and hard bands, respectively.  `V' symbols show
visible $V<20$ stars (Hillenbrand 1997) and `K' symbols indicate
near-infrared $K<18$ stars (Hillenbrand et al.\ 1998, Hillenbrand \&
Carpenter 2000).  The field is centered at $5^h 35^m 21.6^s$ $-5^\circ
22^\prime 12^{\prime\prime}$.  \label{Fig2}}

\begin{figure}
\epsscale{1.0}
\plotone{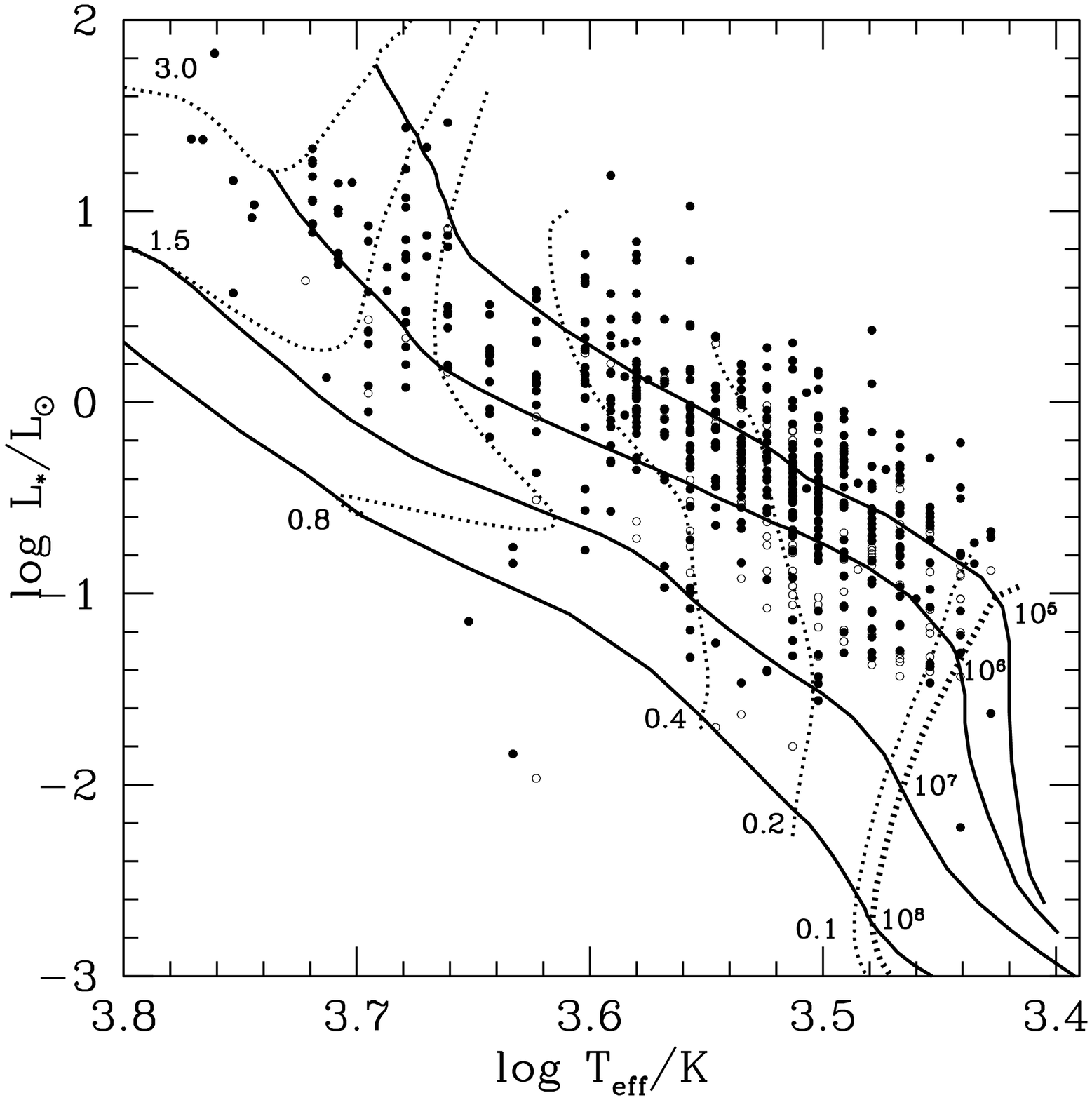}
\end{figure}

\figcaption[garmire.fig3.eps]{Hertzsprung-Russell diagram for
well-characterized ONC stars with $V<20$ in the ACIS field of view
\citep{Hillenbrand97}.  Filled circles indicate detections with ACIS
and open circles indicate nondetections. Evolutionary tracks for
stellar masses $M = 0.1, ..., 3.0$ M$_\odot$ (dotted curves) and
isochrones for ages log$t = 5, ..., 8$ y (solid curves) are from
D'Antona \& Mazzitelli (1997).  Brown dwarfs lie below the dashed line
at $M = 0.08$ M$_\odot$.  \label{Fig3}}
 
\begin{figure}
\epsscale{1.0}
\plotone{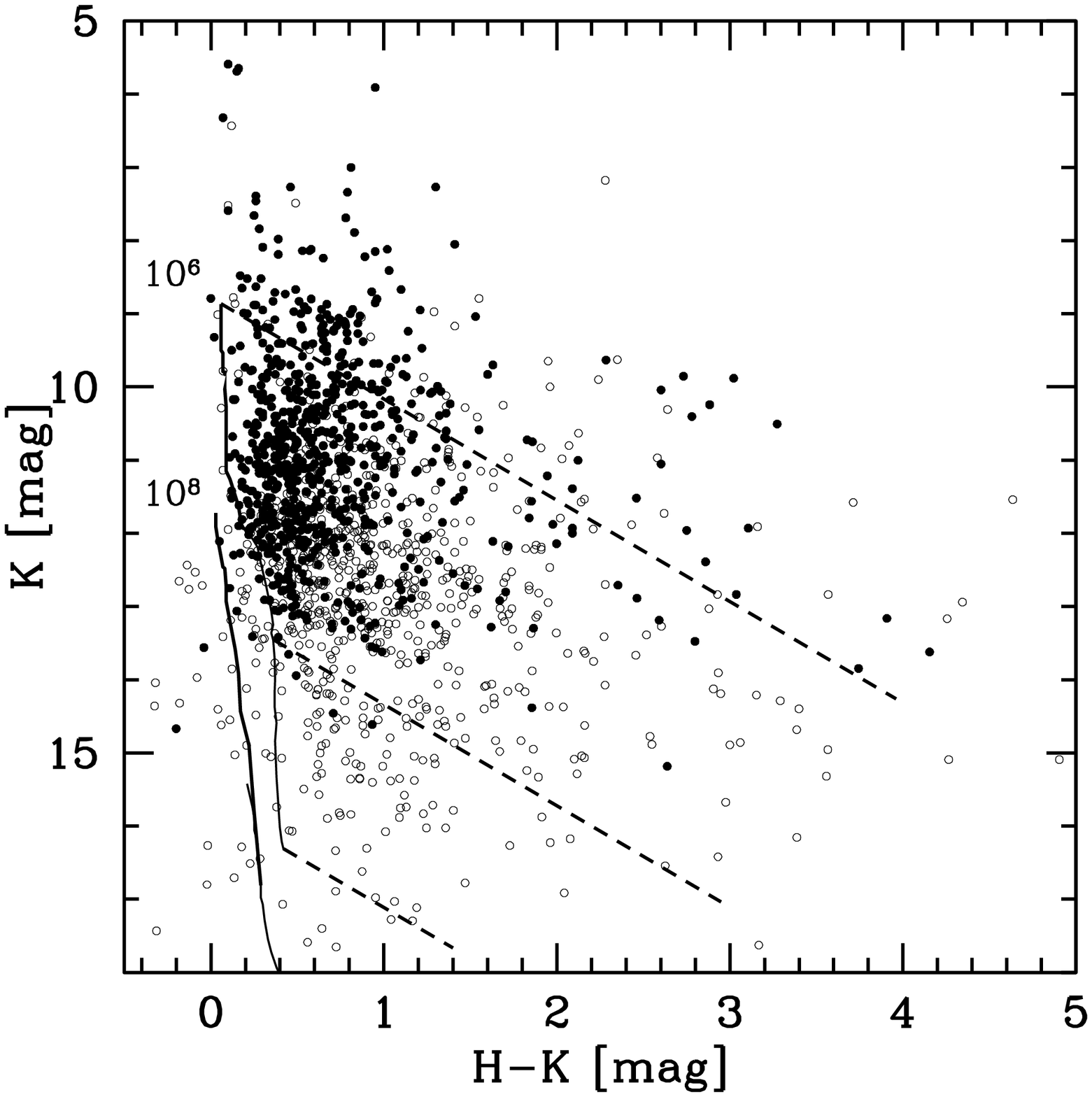}
\end{figure}

\figcaption[garmire.fig4.eps]{Near-infrared color-magnitude diagram for all
objects with $K<18$ in the ACIS field of view \citep{Hillenbrand98,
Hillenbrand00}.  Filled circles are X-ray detections and open circles
are nondetections.  The vertical solid curves show the loci for
unreddened stars with ages log$t = 6, 8$ y.  The dashed lines show
the effects of reddening:  $0 < A_V < 60$ for $M = 2.5$ M$_\odot$
(top); $0 < A_V < 40$ for $M = 0.08$ M$_\odot$ (middle); and $0 < A_V <
15$ for $M = 0.02$ M$_\odot$ (bottom).  \label{Fig4}}

\begin{figure}
\epsscale{0.7}
\plotone{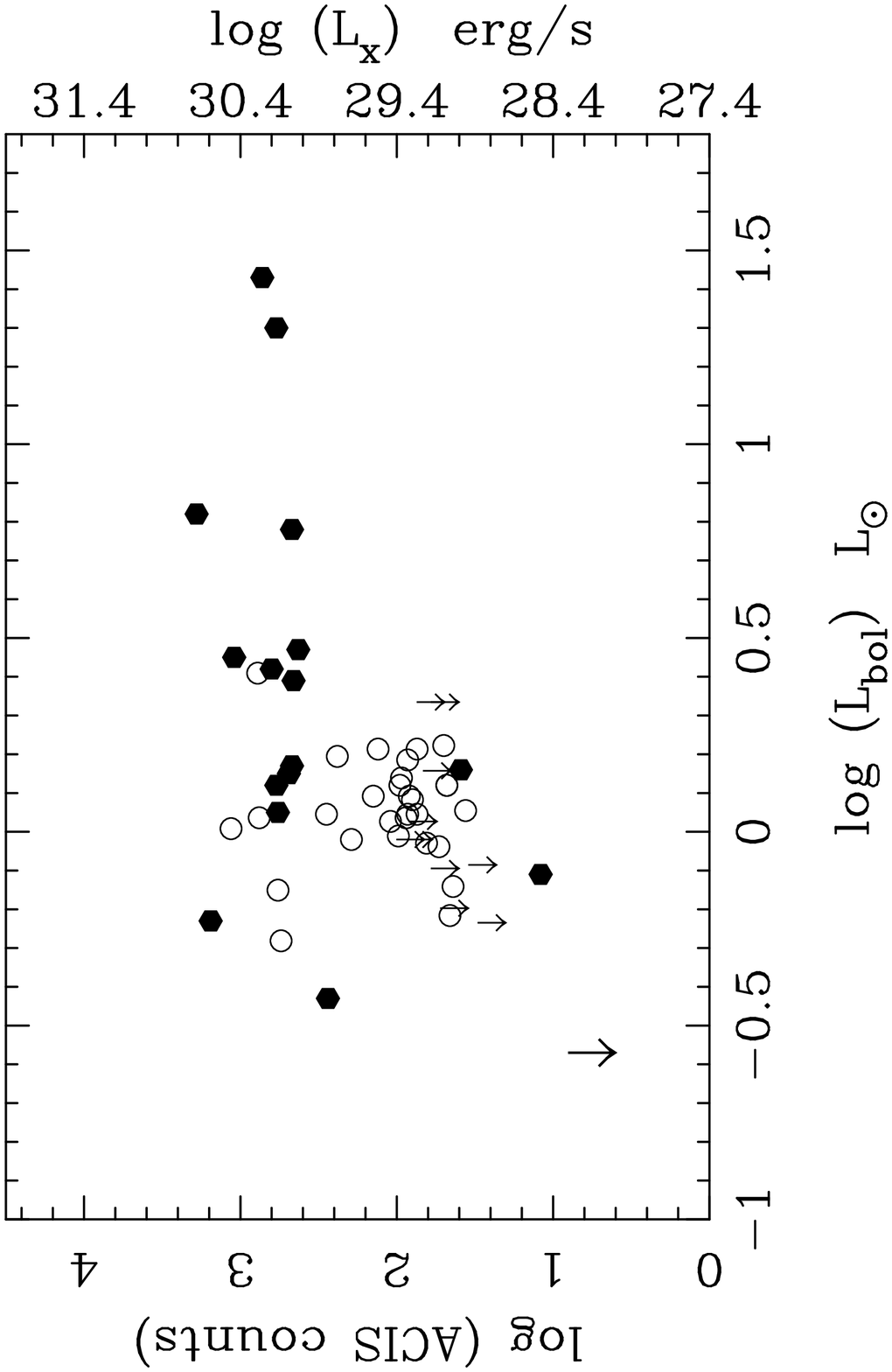}
\end{figure}

\figcaption[garmire.fig5.eps]{Scatter plot of ACIS counts in the soft ($0.2-2$
keV) band and bolometric luminosity for a complete sample of solar-mass
ONC stars with $0.8 < M < 1.2$ M$_\odot$ and $A_V < 3$ (filled circles
and large arrow).  The right axis gives an approximate conversion to soft
X-ray luminosity assuming a $kT = 1$ keV thermal plasma.  Zero-age main
sequence Pleiads in the same mass range are shown for comparison (open
circles and small arrows).  \label{Fig5}}

\begin{figure}
\epsscale{1.0}
\plotone{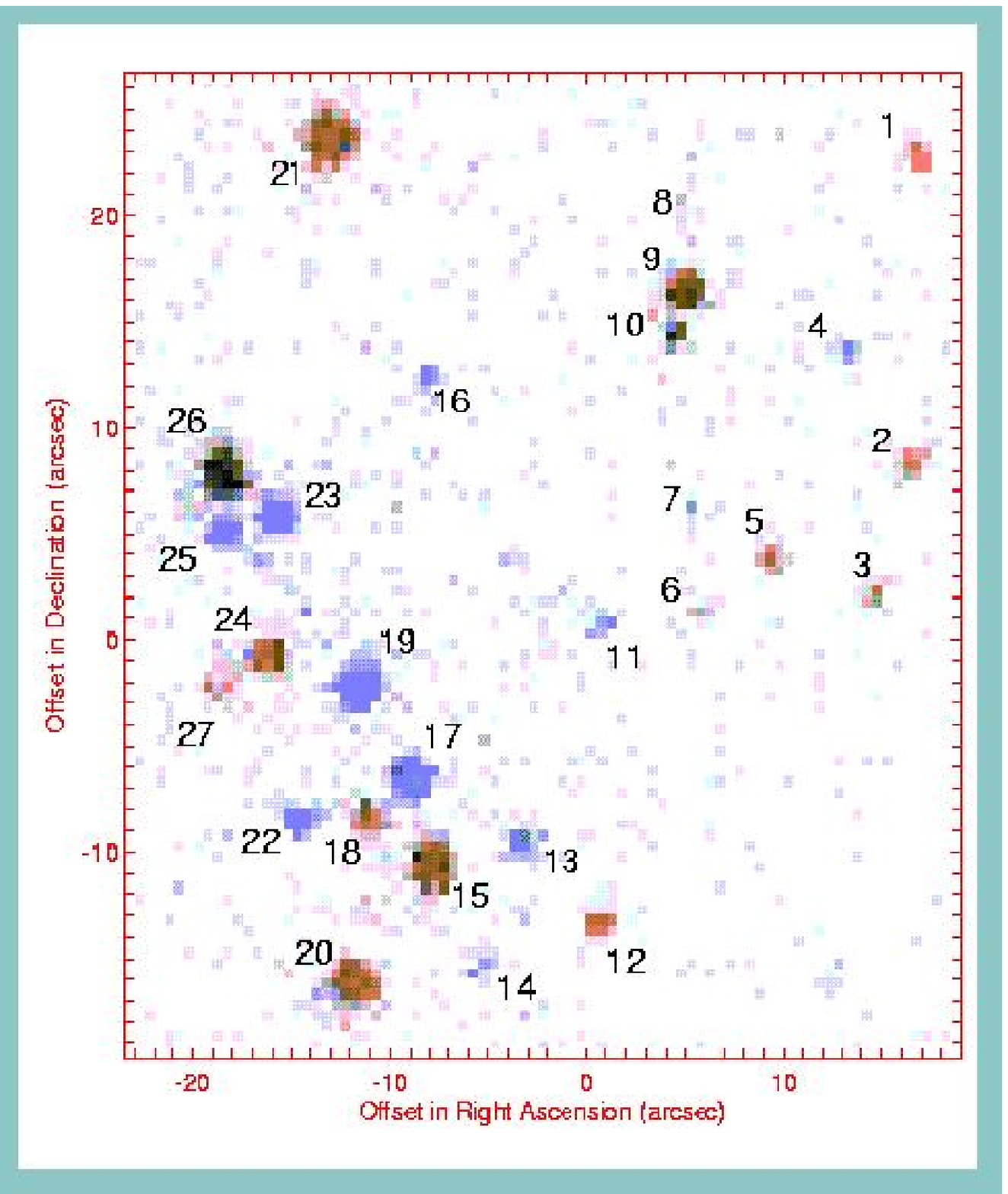}
\end{figure}

\figcaption[garmire.fig6.eps]{Expanded view of the
Becklin-Neugebauer/Kleinman-Low region.  In this figure, color hues are
coded to X-ray energies of individual photons:  red represents photons in
the $0.5-1.5$ keV band, green represent photons in the $1.5-2.5$ keV
band, and blue represents photons in the $2.5-8$ keV band.  Color
intensity is scaled the number of photons in each pixel.  Positions are
indicated with respect to the Becklin-Neugebauer object.  Table 2 gives
source properties and counterparts.  \label{Fig6}}

\newpage

\begin{deluxetable}{rrrrrrr}
\tablewidth{0pt}
\tablecolumns{7}
\tablecaption{ACIS sources in the ONC \label{Table1}}
\tablehead{ \multicolumn{7}{c}{R.A. ~~~ (J2000) ~~~~~ Dec} \\ 
\cline{1-3} \cline{5-7} \\
\colhead{h} & \colhead{m} &\colhead{s} & \colhead{} & 
\colhead{$^\circ$} & \colhead{$\arcmin$} & \colhead{$\arcsec$}
}

\startdata
\cutinhead{Complete sample (831)}
5   &    34   &   40.1  &&  -5  &   26  &   43 \\
5   &    34   &   41.6  &&  -5  &   26  &   51 \\
5   &    34   &   42.6  &&  -5  &   28  &   35 \\
5   &    34   &   44.4  &&  -5  &   24  &   36 \\
\multicolumn{3}{c}{\nodata} && \multicolumn{3}{c}{\nodata} \\
\cutinhead{Additional sources (142)}
5   &    34   &   40.8  &&  -5  &   26  &   35 \\
5   &    34   &   47.5  &&  -5  &   28  &   16 \\
5   &    34   &   48.2  &&  -5  &   16  &   38 \\
5   &    34   &   50.2  &&  -5  &   29  &   13 \\
\multicolumn{3}{c}{\nodata} && \multicolumn{3}{c}{\nodata} \\

\enddata

\tablecomments{The complete version of these tables are in the electronic
edition of the Journal.  The printed edition contains only the first few lines.}

\end{deluxetable}

\newpage

\begin{deluxetable}{rrrrrrrccrcrc}
\tablewidth{0pt}
\tablecolumns{13}
\tabletypesize{\small}
\tablecaption{ACIS Sources Near the Becklin-Neugebauer Object \label{Table2}}
\tablehead{
  \multicolumn{5}{c}{ACIS sources}        & \multicolumn{8}{c}{Counterparts} \\ 
\cline{1-5} \cline{7-13} \\
\colhead{\#}  & \multicolumn{2}{c}{R.A. ~~~ Dec} & \colhead{Counts} & \colhead{log$N_H$} 
   & \colhead{$\theta$} & \colhead{Class} &   \multicolumn{2}{c}{Optical} && \multicolumn{2}{c}{Infrared} & \colhead{Notes} \\
\cline{2-3} \cline{8-9} \cline{11-12} 
\colhead{}   & \colhead{}& \colhead{}& \colhead{} & \colhead{cm$^{-2}$}  
   & \colhead{$^{\prime\prime}$}  & \colhead{} & \colhead{ID} & \colhead{$I$} && \colhead{ID} & \colhead{$K$} & \colhead{} \\
}
 
\startdata
  1 &  5 35 13.03  & -5 22 00.8 &  35~& 21.5  & 0.2   & ONC &JW395  & 15.3  && HC544 & 12.3  & \tablenotemark{a} \\

  2 &  5 35 13.05  & -5 22 15.3 &  40~& 21.5  & 0.1   & ONC &PC 029 & 14.2  && HC505 & 11.0  & \tablenotemark{b} \\

  3 &  5 35 13.17  & -5 22 21.7 &  28~& 22.0  & 0.5   & ONC &JW399ab& 12.8  && HC487 & 10.3  & \tablenotemark{c} \\
  
  4 &  5 35 13.26  & -5 22 10.0 &  16~& 22:~  & 0.2   & IR  &\nodata&\nodata&& HC523 & 12.8 &   \\
       
  5 &  5 35 13.50  & -5 22 19.6 &  36~& 21.5  & 0.3   & ONC &JW411  & 13.5  && HC495 & 10.3  & \tablenotemark{d} \\

  6 &  5 35 13.75  & -5 22 22.4 &  14~&\nodata& 0.4   & ONC &JW420  & 13.6  && HC483 &  9.6  &   \\
      
  7 &  5 35 13.77  & -5 22 17.4 &   9~&\nodata& 0.2   & IR  &\nodata&\nodata&& HC499 & 11.0  & \tablenotemark{e} \\

  8 &  5 35 13.79  & -5 22 03.4 &   8~&\nodata& 0.7   & ONC &JW424  &\nodata&& HC541 & 11.5  &   \\

  9 &  5 35 13.80  & -5 22 07.2 & 365~& 21.5  & 0.1   & ONC &JW423  & 12.3  && HC703 &  9.3  & \tablenotemark{f} \\
  
 10 &  5 35 13.84  & -5 22 09.2 &  59~& 22.0  & 0.2   & IR  &\nodata&\nodata&& HC525 & 11.8  &   \\

 11 &  5 35 14.06  & -5 22 22.2 &  18~& 23.0  & 1.1   & IR  &\nodata&\nodata&& BN    &  5.1  & \tablenotemark{g} \\
   
 12 &  5 35 14.08  & -5 22 37.1 &  53~& 21.5  & 0.5   & ONC &JW432  & 13.0  && HC438 &  9.8  & \tablenotemark{h} \\

 13 &  5 35 14.35  & -5 22 32.9 &  61~& 23.0  & 0.2   & IR  &\nodata&\nodata&&Source n&10.1  & \tablenotemark{i} \\
 
 14 &  5 35 14.50  & -5 22 38.6 &  18~& 23.0  & 0.0   &X-ray&\nodata&\nodata&&\nodata&\nodata& \tablenotemark{j} \\
 
 15 &  5 35 14.66  & -5 22 33.9 & 660~& 21.5  & 0.1   & ONC &JW448  & 12.1  && HC443 &  9.1  & \tablenotemark{k} \\
  
 16 &  5 35 14.67  & -5 22 11.4 &  21~& 23:~  &\nodata&X-ray&\nodata&\nodata&&\nodata&\nodata& \tablenotemark{l} \\

 17 &  5 35 14.73  & -5 22 29.9 & 263~& 22.5  & 0.1   & IR  &\nodata&\nodata&& HC464 & 10.5  & \tablenotemark{m} \\

 18 &  5 35 14.88  & -5 22 31.8 &  97~& 21.5  & 0.2   & ONC &JW452  & 14.6  && HC453 & 11.1  & \tablenotemark{n} \\
 
 19 &  5 35 14.90  & -5 22 25.6 & 426~& 23.0  & 0.1   &X-ray&\nodata&\nodata&&\nodata&\nodata& \tablenotemark{o} \\

 20 &  5 35 14.92  & -5 22 39.4 & 574~& 21.5  & 0.2   & ONC &JW454  & 11.9  && HC431 &\nodata& \tablenotemark{p} \\

 21 &  5 35 14.99  & -5 22 00.1 & 753~& 21.5  & 0.2   & ONC &JW457  & 12.5  && HC546 &  9.8  & \tablenotemark{q} \\
 
 22 &  5 35 15.07  & -5 22 31.3 &  62~& 23.5  & 0.5   & ONC?&PC 086 & 15.6  && HC456 & 12.0  & \tablenotemark{r} \\

 23 &  5 35 15.17  & -5 22 17.7 & 181~& 23.0  &\nodata&X-ray&\nodata&\nodata&&\nodata&\nodata& \tablenotemark{s} \\

 24 &  5 35 15.21  & -5 22 24.2 & 188~& 21.5  & 0.1   & ONC &JW467  & 12.5  && HC478 & 10.3  & \tablenotemark{t} \\

 25 &  5 35 15.35  & -5 22 18.4 &  75~& 23:~  &\nodata&X-ray&\nodata&\nodata&&\nodata&\nodata& \tablenotemark{u} \\
 
 26 &  5 35 15.35  & -5 22 15.8 & 942~& 22.0  & 0.2   & ONC &JW470  & 13.1  && HC504 &  8.1  & \tablenotemark{v} \\
  
 27 &  5 35 15.38  & -5 22 25.7 &  34~& 21.0  & 0.3   & ONC &JW472ab& 12.9  && HC475 & 10.1  & \tablenotemark{w} \\
\enddata  

\tablenotetext{a}{Optical ID: Spectral type M2, $A_V=4.7$, $L=1.1$ L$_\odot$, $M=0.2$ M$_\odot$, log$t < 5$.}
\tablenotetext{b}{ACIS morphology possibly double with separation 1.5\arcsec\/ along P.A. 135$^\circ$. 
    Optical ID = M4, $A_V=0.4$, $L=0.4$ L$_\odot$, $M=0.2$ M$_\odot$, log$t < 5$. }
\tablenotetext{c}{Optical ID = K6.}
\tablenotetext{d}{Optical ID = M1, $A_V=1.3$, $L=0.9$ L$_\odot$, $M=0.3$ M$_\odot$, log$t = 5$.
    Infrared ID = source e.  Radio ID = faint 8.4 GHz radio source.}
\tablenotetext{e}{Infrared ID = source aa.}
\tablenotetext{f}{Optical ID: K2, $A_V=0.0$, $L=1.3$ L$_\odot$, $M=0.2$ M$_\odot$, log$t < 5$. 
    Infrared ID = source g.}

\end{deluxetable}

\newpage

\begin{deluxetable}{r}
\tablenum{1}
\tablecolumns{1}
\tablecaption{}
\tablehead{}
\startdata
~ \\
\enddata

\tablenotetext{g}{Infrared ID = IRc1 = HC705.  Radio ID = radio B. See text for discussion of
possible positional offset.}
\tablenotetext{h}{Optical ID: M3, $A_V=1.8$, $L=2$ L$_\odot$, $M=0.1$ M$_\odot$, log$t < 5$.
    Radio ID: = radio R.}
\tablenotetext{i}{Infrared ID = HC448.  Radio ID = double radio source with 0.4\arcsec separation.}
\tablenotetext{j}{Radio ID = radio H.  The offset is with respect to the radio position. 
    Possible infrared ID = IRc18 (10$\mu$m) lies 1.3\arcsec\/ from the ACIS source \cite{Gezari98}.}
\tablenotetext{k}{Optical ID = K7, $A_V=2.3$, $L=5$ L$_\odot$, $M=0.3$ M$_\odot$, log$t < 5$. 
    Infrared ID = source t.  Radio ID = faint 8.4 GHz radio source.}
\tablenotetext{l}{Radio ID = faint 8.4 GHz radio source.}
\tablenotetext{m}{ACIS count rate drops then rises by factor $\simeq 2$ in $\simeq 5$ hours.
    Optical ID = double with HC465, $V$=10.5.}
\tablenotetext{n}{ACIS count rate rises, falls and rises again over the entire 13 hours
    observation.}
\tablenotetext{o}{ACIS count rate shows slow rise and fall in count rate over 13 hour observation with
    factor $\simeq 2$ amplitude.
    Radio ID = radio D.  The offset is with respect to the radio position}
\tablenotetext{p}{Optical ID = MS-16.  K4, $A_V=2.3$, $L=6$ L$_\odot$, $M=0.8$ M$_\odot$, log$t = 5$.}
\tablenotetext{q}{ACIS count rate drops by factor of $\simeq 2$ in $\simeq 5$ hours. 
    Optical ID = K3, $A_V=2.0$, $L=3$ L$_\odot$, $M=0.8$ M$_\odot$, log$t < 5$.}
\tablenotetext{r}{ACIS count rate decreases after $\simeq 3$ hours. 
    Optical ID = M6, $A_V=0.6$, $L=0.2$ L$_\odot$, $M=0.1$ M$_\odot$, log$t < 5$. 
    See text for discussion of the uncertain counterpart of this X-ray source.}
\tablenotetext{s}{ACIS spectrum is unusually hard out to 7 keV.}
\tablenotetext{t}{Optical ID = K7, $A_V=1.4$, $L=1$ L$_\odot$, $M=0.4$ M$_\odot$, log$t = 5.6$.}
\tablenotetext{u}{ACIS count rate drops by factor of $\simeq 2$ in $\simeq 1$ hour, then
    falls again $\simeq 10$ hours later.
    ACIS spectrum is unusually hard out to 7 keV.}
\tablenotetext{v}{ACIS count rate shows slow decline for $\simeq 10$ hours, followed by rise and fall 
    (flare?) by a factor of $\simeq 2$ within $\simeq 2$ hours. 
    Optical ID = Mid-K}
\tablenotetext{w}{ACIS morphology is possibly double with separation 1.5\arcsec\/ along P.A. 120$^\circ$.}

\end{deluxetable}

\end{document}